\documentclass[aps,prb,twocolumn,superscriptaddress,showpacs]{revtex4-1}

\usepackage{graphicx}
\usepackage{calc}
\usepackage{bm}
\usepackage{color}

\begin{document}

\title{Proximity-induced superconductivity within the InAs/GaSb  edge conducting state.}

\author{A.~Kononov}
\affiliation{Institute of Solid State Physics RAS, 142432 Chernogolovka, Russia}
\author{V.A.~Kostarev}
\affiliation{Institute of Solid State Physics RAS, 142432 Chernogolovka, Russia}
\author{B.R.~Semyagin}
\affiliation{Institute of Semiconductor Physics, Novosibirsk 630090, Russia}
\author{V.V.~Preobrazhenskii}
\affiliation{Institute of Semiconductor Physics, Novosibirsk 630090, Russia}
\author{M.A.~Putyato}
\affiliation{Institute of Semiconductor Physics, Novosibirsk 630090, Russia}
\author{E.A.~Emelyanov}
\affiliation{Institute of Semiconductor Physics, Novosibirsk 630090, Russia}
\author{E.V.~Deviatov}
\affiliation{Institute of Solid State Physics RAS, 142432 Chernogolovka, Russia}

\date{\today}

\begin{abstract}
We experimentally investigate Andreev transport through the interface between an indium superconductor and the edge of the InAs/GaSb bilayer. To cover all possible regimes of InAs/GaSb spectrum, we study samples with  10-nm, 12~nm, and 14~nm thick InAs quantum wells. For the trivial case of a direct band insulator in 10~nm samples, differential resistance demonstrates standard Andreev reflection. For InAs/GaSb structures with band inversion (12~nm and 14~nm samples), we observe distinct low-energy structures, which we regard as  direct evidence for the proximity-induced superconductivity within the  current-carrying edge state. For 14~nm InAs well samples, we additionally observe  mesoscopic-like resistance fluctuations, which are subjected to threshold suppression in low magnetic fields.
\end{abstract}

\pacs{73.40.Qv  71.30.+h}

\maketitle

\section{Introduction}

Similarly to HgTe quantum wells~\cite{konig,kvon}, InAs/GaSb bilayers can demonstrate inverted energy spectra~\cite{hughes08}. For a typical value of 10~nm for the GaSb layer, InAs/GaSb structures with 12~nm thick InAs wells are usually regarded as topological insulators~\cite{hughes08,gasb2,gasb4,nowack14,gasb6,suzuki}. Thinner (10~nm) or thicker (14~nm) InAs wells correspond~\cite{suzuki,tiemann} to a direct band  semiconductor or an indirect band two-dimensional semimetal, respectively. InAs/GaSb bilayers posses many advantages over HgTe quantum wells, including better stability, much easier III-V materials processing and spectra tunability by front and back gates~\cite{hughes08}. However, there  is  residual bulk conductivity in InAs/GaSb structures, which complicates experimental investigation of  edge transport~\cite{suzuki,knez15}.

Topological edge states with spin-momentum locking are expected for structures with band inversion~\cite{zhang1,kane,zhang2,ando}. Current-carrying edge states were demonstrated for InAs/GaSb bilayers in transport experiments~\cite{suzuki,knez15,nowack14,pribiag15,nichele16}, although their topological nature is still debatable~\cite{nichele16}.

Edge state transport is of special interest for regions with proximity-induced superconductivity~\cite{Fu,yakoby,beenakker13}, because of a search for Majorana fermions with non-Abelian statistics~\cite{reviews} and prospects for quantum computing~\cite{beenakker13, kitaev}. This activity requires detailed investigation of Andreev transport in systems with non-trivial energy spectra~\cite{adroguer,visani,finck}. 

\begin{figure}
\includegraphics[width=0.8\columnwidth]{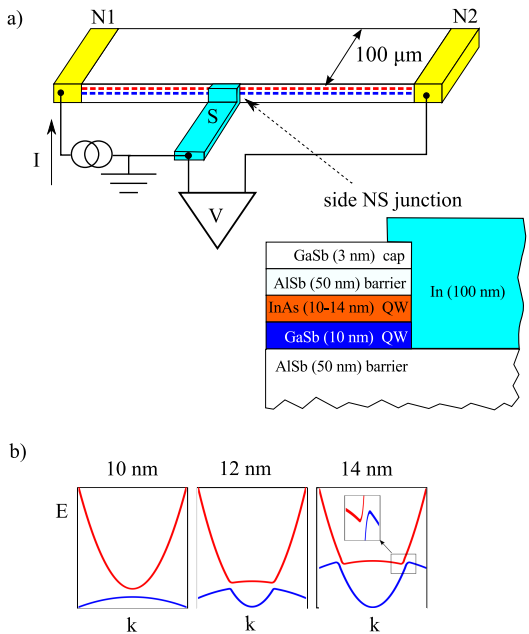}
\caption{(Color online) (a) Sketch of the sample (not to scale) with electrical connections.  10~$\mu$m wide side normal-superconductor In--InAs/GaSb junctions   are fabricated   by lift-off technique, after thermal evaporation of a thick In film (gray) over the mesa step. Charge transport is investigated across a single In-InAs/GaSb junction in a standard three-point technique: the superconducting electrode  is grounded, while Ohmic contacts N1 and N2 (yellow) are employed to feed the current and measure the voltage drop, respectively. (b) Schematic diagrams of the expected energy spectrum for different InAs quantum well thicknesses, see Refs.~\protect\onlinecite{suzuki,tiemann} for details. 
}
\label{sample}
\end{figure}

Andreev reflection~\cite{andreev} allows charge transport from normal metal (N) to superconductor (S) at energies below the superconducting gap. An electron is injected through the NS interface by creating  a Cooper pair, so a hole is reflected back to the N side of the junction~\cite{andreev,tinkham}. Usually, Andreev reflection is not sensitive to  the details of band structure in the normal lead~\cite{tinkham}. However, for graphene or semimetal spectra,  the reflected hole can appear in the valence band, which is known as specular (or interband) Andreev reflection~\cite{been1,been2,spec}. Also, an additional energy scale appears if Andreev transport goes through an intermediate conductive region, which is partially decoupled from the bulk normal conductor~\cite{heslinga,klapwijk17}.

Here, we experimentally investigate Andreev transport through the interface between an indium superconductor and the edge of the InAs/GaSb bilayer. To cover all possible regimes of InAs/GaSb spectrum, we study samples with  10-nm, 12~nm, and 14~nm thick InAs quantum wells. For the trivial case of a direct band insulator in 10~nm samples, differential resistance demonstrates standard Andreev reflection. For InAs/GaSb structures with band inversion (12~nm and 14~nm samples), we observe distinct low-energy structures, which we regard as  direct evidence for the proximity-induced superconductivity within the  current-carrying edge state. For 14~nm InAs well samples, we additionally observe  mesoscopic-like resistance fluctuations, which are subjected to threshold suppression in low magnetic fields.

\section{Samples and technique}

Our samples are grown by solid source molecular beam epitaxy on semi-insulating GaAs (100) substrate. The InAs/GaSb double quantum well is sandwiched between two 50~nm thick AlSb barriers.  Details on the growth parameters can be found elsewhere~\cite{growth}. To cover all possible regimes of the InAs/GaSb bilayer spectrum~\cite{suzuki,tiemann}, we prepare samples with a 10-nm thick GaSb quantum well and different,  10-nm, 12~nm, and 14~nm thick InAs ones, see Fig.~\ref{sample}. 

As obtained from standard magnetoresistance measurements, the 10~nm and 14~nm samples are characterized by bulk electron-type conductivity, while it is hole-type for the 12~nm ones. The low-temperature mobility is found to be one order of magnitude higher for bulk  electrons ($ 10^{4}  $~cm$^{2}$/Vs) than  for holes ($ 10^{3}  $~cm$^{2}$/Vs). These values are in good correspondence with known ones for InAs/GaSb double quantum wells~\cite{hughes08,gasb2,gasb4,nowack14,gasb6}, taking into account low bulk carrier concentration, which is roughly $\approx 4 \cdot 10^{11}  $cm$^{-2}$  in all our samples.

A sample sketch is presented in Fig.~\ref{sample} (a).  The $80$~nm high mesa is formed by wet chemical etching down to the bottom GaSb layer. Since the edge effects are of prime interest in InAs/GaSb bilayers~\cite{hughes08,gasb2,gasb4,nowack14,gasb6}, side~\cite{nbsemi,nbhgte} superconducting contacts are made at the mesa step. They are formed from thermally evaporated 100~nm thick indium film by lift-off with low (1-2~$\mu$m) mesa overlap, see Fig.~\ref{sample} (a). Because of the insulating top AlSb barrier, vertical transport is forbidden in the overlap region. We take special care to obtain equally prepared In--InAs/GaSb interfaces. Both the processing steps, wet etching and indium evaporation, are made simultaneously for samples with  10-nm, 12~nm, and 14~nm thick InAs quantum wells.  Ohmic contacts are made by thermal evaporation of 100~nm Au film with few nm Ni to improve adhesion.

We study charge transport across a single NS junction between the indium side contact and the InAs/GaSb mesa edge in a standard three-point technique, see Fig.~\ref{sample} (a): the superconducting electrode  is grounded; a current (-2 to +2~$\mu$A range) is fed to InAs/GaSb bilayer through one of the normal Ohmic contacts, N1 in Fig.~\ref{sample}; the other normal contact (N2, respectively) traces the potential $V$. 

 To obtain $dV/dI(V)$ characteristics,   the current  is additionally modulated by a low ac (20~pA, 110~Hz) component. We measure both,  dc ($V$) and ac ($\sim dV/dI$), components of the  potential by using a dc voltmeter and a lock-in, respectively. We check, that the lock-in signal is independent of the modulation frequency in the 60~Hz -- 300~Hz range, which is defined by applied ac filters.  To extract features specific to the InAs/GaSb bilayer system, the measurements are performed at  30~mK. Similar results are obtained from different samples in several cooling cycles.

\begin{figure}
\includegraphics[width=\columnwidth, trim = 0mm 0mm 0mm 0mm, clip=true]{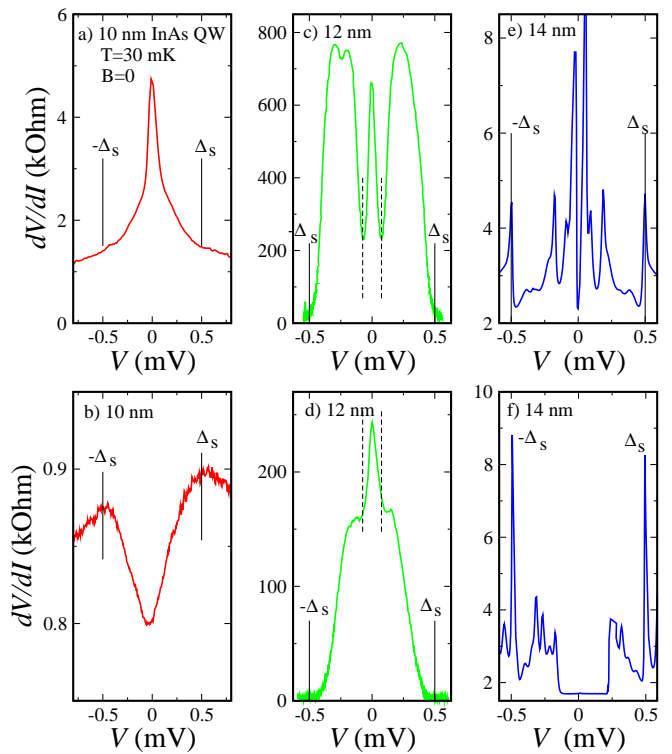}
\caption{(Color online) $dV/dI(V)$ curves for a single In--InAs/GaSb junction for different samples. For every curve, $dV/dI$ is finite within the indium superconducting gap $|eV|<\Delta_s=0.5$~meV due to Andreev reflection. The top and the bottom panels demonstrate maximum device-to-device fluctuations for a given InAs quantum well thickness: (a-b) 10~nm, which is expected to have trivial insulator band structure. There is no any additional $dV/dI$ features; (c-d) 12~nm, a supposed topological insulator. There is well-developed $dV/dI$ peak within $\pm 0.07$~mV, the subgap $dV/dI$ resistance is extremely high, about $200-800$~kOhm; (e-f) 14~nm, two-dimensional indirect-band semimetal. A zero-bias resistance dip  is accompanied by a number of additional symmetric peaks of different amplitude. The curves are obtained at 30~mK in zero magnetic field.}
\label{IV}
\end{figure}

\section{Experimental results}

Fig.~\ref{IV}  demonstrates  examples  of $dV/dI(V)$ curves for samples with different thickness of InAs quantum well.  In a three-point technique, the measured potential $V$ reflects in-series connected resistances of the grounded contact and some part of the 2D system. In our experiment the former term is dominant, because of highly resistive junctions, see Fig.~\ref{IV}. The indium lead is superconducting, so $dV/dI(V)$ characteristics  reflect  charge transport through a single (grounded) NS interface.  To support this conclusion experimentally, the obtained $I-V$ characteristics are verified to be independent of the exact positions of  current and voltage probes.

Despite of equally prepared In--InAs/GaSb interfaces, $dV/dI(V)$ curves demonstrate even qualitative different behavior in samples with different thickness of InAs quantum well in Fig.~\ref{IV}. Since the $dV/dI(V)$ curves of NS junctions are known to be highly sensitive to the interface potential fluctuations~\cite{BTK}, the top and the bottom panels in Fig.~\ref{IV} demonstrate maximum device-to-device fluctuations for a given InAs quantum well thickness.

The 10~nm wide InAs quantum well sample demonstrates  a typical example of Andreev reflection at the disordered NS interface~\cite{tinkham}, see Fig.~\ref{IV}~(a-b). In (a), the differential resistance $dV/dI$ is increased within the indium superconducting gap $|eV|<\Delta_s=0.5$~meV to about $5~k\Omega$, so single-particle scattering is significant at the interface~\cite{BTK}. In Fig.~\ref{IV}~(b), there is a resistance drop within $|eV|<\Delta_s=0.5$~meV, as it is expected for cleaner NS interface~\cite{andreev,BTK}. There is no any additional $dV/dI$ features for the curves in Fig.~\ref{IV}~(a-b), as it should be anticipated for standard Andreev reflection.

\begin{figure}
\includegraphics[width=\columnwidth, trim = 0mm 0mm 0mm 0mm, clip=true]{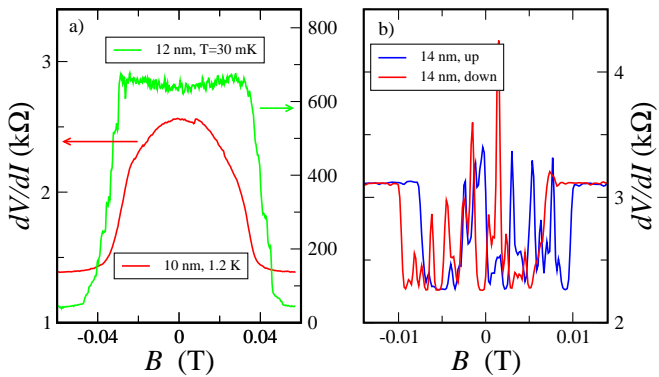}
\caption{(Color online) (a) Suppression of the superconductivity by in-plane magnetic field for the 10~nm sample at 1.2~K and for the  12~nm one at 30~mK. The resistance drop is clearly broadened at high temperature. Monotonous $dV/dI(B)$ suppression is fully consistent with the classical  Andreev reflection picture~\protect\cite{tinkham}.
(b) Threshold suppression of the mesoscopic-like resistance fluctuations  by low in-plane magnetic field for the 14~nm sample at 30~mK.  The exact threshold positions depend slightly on the magnetic field direction.  The dc bias is fixed at $V=0$ during the field sweep.
}
\label{RB}
\end{figure}

In Fig.~\ref{IV}~(c-d), the subgap  $dV/dI$ resistance is about $200-800$~kOhm in different samples at $|eV|<\Delta_s=0.5$~meV. It is much higher than the normal $dV/dI$ value $\approx 20$~kOhm for $|eV|>\Delta_s$. Which is specific for the 12~nm samples, there is always well-developed $dV/dI$ peak within $\pm 0.07$~mV bias. 

\begin{figure}
\includegraphics[width=\columnwidth, trim = 0mm 0mm 0mm 0mm, clip=true]{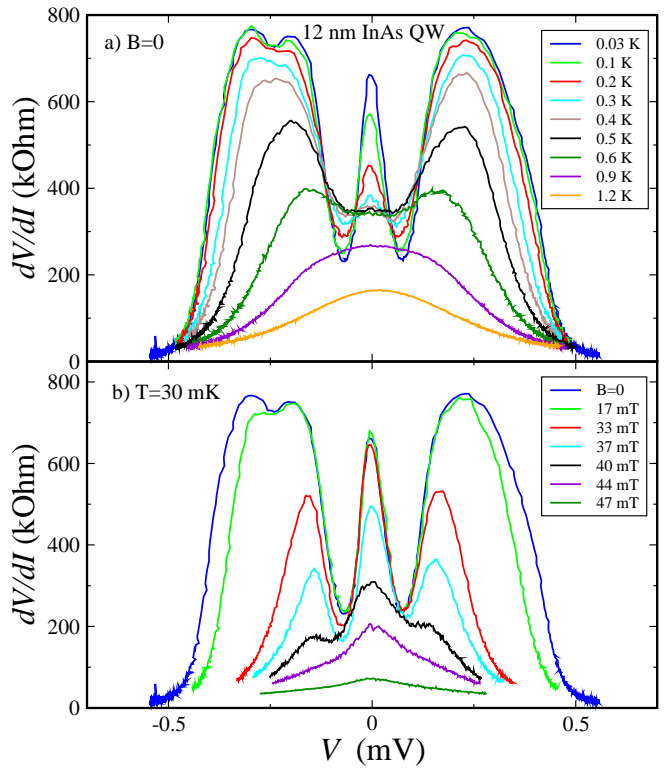}
\caption{(Color online) $dV/dI(V)$ behavior with temperature (a) or magnetic field (b) for the 12~nm sample. The superconductivity is gradually suppressed, but the resistance peak within $\pm 0.07$~mV is well visible for temperatures below $0.6~K$ in zero field (a) and below 40~mT at  minimal temperature (b). 
}
\label{12nmIV}
\end{figure}

The $dV/dI(V)$ behavior is even more complicated for the 14~nm samples, see  Fig.~\ref{IV}~(e-f). Differential resistance demonstrates sharp peaks at $|eV|=\Delta_s$, the subgap and normal $dV/dI$ values are comparable with the 10~nm case. However,  a zero-bias resistance dip  appears in Fig.~\ref{IV}~(e-f), which is accompanied by a number of additional resistance peaks of different amplitude. They are well reproducible for a given sample and symmetric in respect to the bias sign.

First of all, we demonstrate that the presented in Fig.~\ref{IV} $dV/dI(V)$ curves are connected with superconductivity.  Fig.~\ref{RB} (a) demonstrates that the superconductivity can be completely suppressed above $\approx 30~mT$, which well corresponds to the known~\cite{indium} bulk indium critical filed. We fix the zero bias $V=0$ and  sweep the magnetic field slowly. The resistance drop is sharp at 30~mK (as shown for the 12 nm sample), while temperature broadening is demonstrated  at 1.2~K for the 10~nm sample in Fig.~\ref{RB}.  To avoid orbital effects, the field is oriented within the bilayer plane (with $0.5^\circ$ accuracy) along the mesa edge, so it is strictly in-plane oriented also for the superconducting film at the mesa step. We obtain similar  results  for the normally oriented magnetic field. Monotonous $dV/dI(B)$ suppression is fully consistent with the classical  Andreev reflection picture~\cite{tinkham}.

Fig.~\ref{RB} (b) demonstrates specifics of the low-field behavior for the 14~nm samples.  One can see  strong mesoscopic-like $dV/dI(B)$ fluctuations within $\approx \pm 10 mT$ interval, which are completely suppressed at higher fields. 

Fig.~\ref{12nmIV} demonstrates detailed $dV/dI(V)$ behavior with temperature or magnetic field increase for the 12~nm sample. The resistance peak within $\pm 0.07$~mV is well visible for temperatures below $0.6$~K, see Fig.~\ref{12nmIV} (a), and for magnetic fields below 40~mT in Fig.~\ref{12nmIV}~(b). At the temperature of $1.2~K$, $dV/dI(V)$ curve is still nonlinear because of much higher indium  $T_c\approx 3.4~K$.  The superconductivity is gradually suppressed above 30~mT in Fig.~\ref{12nmIV}~(b), further increase of magnetic field results in a nearly flat curve even at lowest $T=30$~mK.

\section{Discussion} \label{disc}

Within the classical framework of  Andreev reflection~\cite{tinkham}, it is not sensitive to  details of band structure  in a normal lead. However, even qualitative effect on $dV/dI(V)$ can be seen in Fig.~\ref{IV} for samples with different InAs quantum well widths. Because the 
observed subgap features are independent of the maximum device-to-device fluctuations, we have to attribute them  to different edge properties of our InAs/GaSb structures, which are defined by bulk spectrum~\cite{hughes08,gasb2,gasb4,nowack14,gasb6}.

No edge specifics can be expected for a trivial insulator in 10~nm thick InAs quantum well samples.  Monotonous $dV/dI(V)$ curves in Fig.~\ref{IV} (a-b) do not demonstrate subgap features, they are only sensitive to the disorder at the interface~\cite{BTK}.

In the case of 12~nm thick InAs quantum well, the current-carrying edge states appear, because of the inverted band structure. This statement seems to be firmly confirmed by experiments~\cite{hughes08,gasb2,gasb4,nowack14,gasb6}. Moreover, the edge current was directly demonstrated in visualization experiments~\cite{nowack14,imaging} to  coexist with finite bulk conductivity, most likely due to the edge depletion region. The latter leads to strongly increased differential resistance in Fig.~\ref{IV} (c-d).   The proximity-induced superconducting gap $\Delta_{ind}$ can be expected within the  edge state near the indium superconducting lead~\cite{adroguer}. 

\begin{figure}
\includegraphics[width=\columnwidth, trim = 0mm 0mm 0mm 0mm, clip=true]{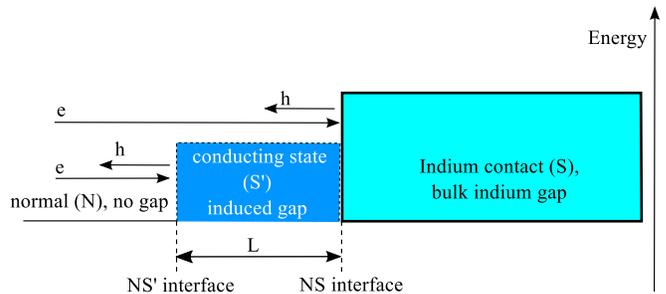}
\caption{(Color online) Schematic energy diagram of the edge region for InAs/GaSb structures with band inversion. The proximity-induced superconductivity (blue region S') can be expected within the conductive edge state  near the indium superconducting lead~\protect\cite{adroguer}. The NS' interface is responsible for Andreev reflection at  biases below the induced gap $|eV|<\Delta_{ind}$, while above this value the NS interface with bulk superconductor (S) governs the reflection process. Because of different single-particle transparency of the interfaces, $dV/dI(V)$ contains~\protect\cite{klapwijk17} an additional structure at low biases in Fig.~\protect\ref{IV} (c-d) and (e-f)}
\label{discussion}
\end{figure}

Andreev transport through the intermediate conductive region has been regarded both experimentally~\cite{heslinga,klapwijk17} and theoretically~\cite{akhmerov17}. In a crude qualitative picture, see Fig.~\ref{discussion}, the NS' interface with the region of induced superconductivity S' is responsible for Andreev reflection at  biases below the induced gap $|eV|<\Delta_{ind}$, while above this value the NS interface with bulk superconductor governs the reflection process. Because of different single-particle transparency of two interfaces, $dV/dI(V)$ contains~\cite{heslinga,klapwijk17} an additional structure at low biases. The induced gap $\Delta_{ind}$ can be estimated from the width of this structure~\cite{heslinga,klapwijk17} in Fig.~\ref{IV} (c-d) as $0.07$~meV.

Because the edge conductive region is of finite width~\cite{nowack14,imaging} $L$,  the induced gap should be defined by Thouless energy $\Delta_{ind} \sim E_{Th}$ (see Appendix to Ref.~\onlinecite{akhmerov17} for recent comprehensive discussion). This statement is in qualitative agreement with our experiment: (i) as expected~\cite{golubov15} for $E_{Th}$,  the width of the low-bias structure is constant in Fig.~\ref{12nmIV} (a), until $k_BT$  exceeds $\Delta_{ind}$ at 0.9~K; (ii) also, $E_{Th}$ is insensitive~\cite{golubov15} to partial suppression of the bulk superconducting gap by magnetic field, as we observe in Fig.~\ref{12nmIV}~(b). As for numerical estimations,  $ E_{Th}=\hbar D/L^2$ in the regime of diffusive transport. If we use the bulk values,  $v_F\approx6\times10^{4}~m/s$ and $l\approx 10$~nm, we can estimate $L$ as 60~nm from the experimental value of $E_{Th}\approx 0.07$~meV.  This  crude estimation corresponds well to the experimentally obtained~\cite{pribiag15} value  $L<260$~nm.

Because of the band inversion, we can also expect~\cite{tiemann} the edge conductive region for samples with 14~nm width of the InAs quantum well. Thus, the zero-bias resistance dip in Fig.~\ref{IV}~(e-f) can also be regarded as the induced gap $\Delta_{ind}\sim E_{Th}$. However, the subgap resistance peaks in Fig.~\ref{IV}~(e-f) and  mesoscopic-like fluctuations in low fields in Fig.~\ref{RB} (b) resemble  modulation~\cite{akhmerov17,chevallier} of density of states due to the quasiparticle interference~\cite{akhmerov17}. It appears~\cite{adroguer,tomasch1} for ballistic $l>>L$ transport, which seems to be reasonable for $l\approx 100$~nm in the 14~nm samples. In this case, the threshold suppression of the mesoscopic-like fluctuations reflects the interference breakdown in magnetic field. It is important, that we do not observe any subgap features for the 10~nm samples with the similar $l\approx 100$~nm value, where no edge conductive region can be expected. Thus, we can regard the subgap resistance features in Fig.~\ref{IV} (c-d) and (e-f) as a direct evidence for the proximity-induced superconductivity within the  current-carrying edge states in  InAs/GaSb structures with band inversion.

\section{Conclusion}

As a conclusion, we experimentally investigate Andreev transport through the interface between an indium superconductor and the edge of the InAs/GaSb bilayer. To cover all possible regimes of InAs/GaSb spectrum, we study samples with  10-nm, 12~nm, and 14~nm thick InAs quantum wells. For the trivial case of a direct band insulator in 10~nm samples, differential resistance demonstrates standard Andreev reflection. For InAs/GaSb structures with band inversion (12~nm and 14~nm samples), we observe distinct low-energy structures, which we regard as  direct evidence for the proximity-induced superconductivity within the  current-carrying edge state. For 14~nm InAs well samples, we additionally observe  mesoscopic-like resistance fluctuations, which are subjected to threshold suppression in low magnetic fields.

\acknowledgments

We wish to thank Ya.~Fominov, D.E.~Feldman, V.T.~Dolgopolov, and T.M.~Klapwijk for fruitful discussions. We gratefully acknowledge financial support by the RFBR (project No.~16-02-00405) and RAS.

\end{document}